\documentstyle[12pt,a41,epsfig]{article}

\newcommand{\beq}{\begin{equation}}
\newcommand{\eeq}{\end{equation}}

\newcommand{\gsim}{\stackrel{>}{\sim}}
\newcommand{\lsim}{\stackrel{<}{\sim}}

\newcommand\MeV{\,\mbox{MeV}}
\newcommand\GeV{\,\mbox{GeV}}

\newcommand\pvec{\mbox{\boldmath $p$}}

\newcommand{\ds}{\displaystyle}
\newcommand{\la}[1]{\label{#1}}
\newcommand{\re}[1]{\ (\ref{#1})}
\newcommand{\nn}{\nonumber}
\newcommand{\be}{\begin{equation}}
\newcommand{\ee}{\end{equation}}
\newcommand{\ba}{\begin{eqnarray}}
\newcommand{\ea}{\end{eqnarray}}
\newcommand{\baz}{\begin{eqnarray*}}
\newcommand{\eaz}{\end{eqnarray*}}
\newcommand{\ct}[1]{${\cite{#1}}$}
\newcommand{\ctt}[2]{${\cite{#1}-\cite{#2}}$}
\newcommand{\bi}[1]{\bibitem{#1}}
\newcommand{\ep}{\epsilon}
\newcommand{\epm}{\varepsilon_{\mu\nu\lambda\sigma}}
\newcommand{\epa}{\varepsilon_{\mu\alpha\nu\beta}}
\newcommand{\Sa}{S_{\mu\alpha\nu\beta}}
\newcommand{\gb}{\gamma_\beta}
\newcommand{\g}{\gamma_5}
\newcommand{\lc}{\lambda^c}
\newcommand{\bq}{\bar q}

\begin{document}
\sloppy
\thispagestyle{empty}
\begin{flushleft}
DESY 96--040 \\
March 1996\\
\end{flushleft}

\setcounter{page}{0}

\mbox{}
\vspace*{\fill}
\begin{center}
{\LARGE\bf On the Twist-2 Contributions to
} \\
\vspace{2.5mm}
{\LARGE\bf Polarized Structure Functions
}\\
\vspace{2.5mm}
{\LARGE\bf and New Sum Rules
}\\
\vspace{4em}
\large
J. Bl\"umlein$^a$ and N. Kochelev$^{a,b}$

\vspace{4em}
\normalsize
{\it   $^a$DESY--Zeuthen}\\
{\it   Platanenallee 6, D--15735 Zeuthen, Germany}\\
\vspace{5mm}
{\it   $^b$Bogoliubov
Laboratory of Theoretical Physics, JINR,
}\\
{\it   RU--141980 Dubna, Moscow Region,
Russia}\\
\end{center}
\normalsize
\vspace*{\fill}
\begin{abstract}
\noindent
The twist-2 contributions to the polarized structure functions
in deep inelastic lepton--hadron scattering are calculated including
the exchange of weak bosons and using both
the operator product expansion and the covariant parton model.
A new relation between two structure functions leading to a
sequence of new sum rules is found.
The light quark mass corrections to the structure functions are derived
in lowest order QCD.
\end{abstract}
\vspace*{\fill}
\newpage
%
\section{Introduction}
\label{sect1}
%
The study of polarized deep inelastic scattering off polarized targets
has revealed a rich structure of phenomena during the last
years~\cite{REV}. So far mainly the case of deep inelastic photon
scattering has been studied experimentally. A future polarized proton
option at RHIC and HERA, however, would allow to probe the spin
structure of nucleons at much higher $Q^2$~(cf.~\cite{JB95A}) also.
In this
range $Z$--exchange contributions become relevant and one may investigate
charged current scattering as well.
For this general case
the scattering cross section is  determined by (up to) five polarized
structure
functions per current combination, if lepton mass effects are
disregarded.

In previous investigations different techniques have been used to derive
relations between these structure functions and discrepancies between
several derivations were reported~(cf. e.g.~\cite{a1}).
In refs.~\cite{a1,a0} the structure functions were calculated in
the parton model. Some of the investigations deal with the
case of longitudinal polarization only~\cite{a3}.
In other studies light-cone current algebra~\cite{DIC,A1C} and
the operator product expansion were
used~\cite{AR}--\cite{a77}. Furthermore  the structure
functions $g_1^{em}$
and $g_2^{em}$ were also
calculated in the covariant parton model~\cite{a1B,a17}.
Still a thorough agreement between  different approaches has not
been obtained.

It is the aim of
the present paper to derive
the relations for the complete
set of the polarized structure functions including weak interactions
which are not associated with terms in the
scattering cross section
vanishing as $m_{lepton} \rightarrow 0$.
The calculation is performed applying two different
techniques:~the operator product
expansion and  the covariant parton model~\cite{LP}.
The latter method is
furthermore used
to obtain also
the quark mass corrections in lowest
order QCD.

As it turns out the twist-2 contributions for
only
two out of the five polarized structure
functions, corresponding to the respective current combinations, are
linearly independent. Therefore three
linear operators have to exist which
determine the remaining three structure
functions over a basis of two in lowest order QCD.
Two of them are given by the
Wandzura--Wilczek\cite{WW} relation  and a
relation by Dicus\footnote{This relation corresponds
to the
Callan--Gross~\cite{CG}
relation for unpolarized structure functions since
the spin dependence enters the tensors of $g_4$ and $g_5$ in
$W_{\mu\nu}^{ij}$,~eq.~(\ref{eqz4}), in terms of a factor
$S.q$.}~\cite{DIC}.
A third {\it new}
relation is found.

New sum rules based on this relation are derived and  discussed
in the context of quark mass corrections. Extending a recent analysis
carried out for the case of photon scattering~\cite{a18}
to the complete set
of neutral and charged current interactions
we also investigate the validity
of known relations, as the Burkhardt--Cottingham~\cite{BC}
sum rule and other
relations, in the presence of quark mass effects.

%
\section{Basic Notation}
\label{sect2}
%
The hadronic tensor for polarized deep inelastic scattering is given by
\be
W_{\mu\nu}^{ab}=\frac{1}{4\pi}\int d^4xe^{iqx}
\langle pS\mid[J_\mu^a(x),J_\nu^b(0)]\mid pS\rangle,
\label{eqHAD}
\ee
where in framework of the quark model the currents are
\be
J_\mu^a(x)=\sum_{f,f'}
U_{ff'} \overline{q}_{f'}(x)\gamma_\mu(g_V^a+g_A^a\gamma_5)q_f(x).
\ee
In terms of structure functions the hadronic tensor reads:
\ba
W_{\mu\nu}^{ab}
&=&(-g_{\mu\nu}+\frac{q_\mu q_\nu}{q^2})F_1^i(x,Q^2)+
\frac{\widehat{p}_\mu\widehat{p}_\nu}{p.q} F_2^i(x,Q^2)-
 i\epm\frac{q_\lambda p_\sigma}{2 p.q}  F_3^i(x,Q^2)\nn\\
&{+}& i\epm\frac{q^\lambda S^\sigma}{p.q} g_1^i(x,Q^2)+
i\epm\frac{q^{\lambda}(p.q S^\sigma - S.q p^\sigma)}
{(p.q)^2} g_2^i(x,Q^2)\nn\\
&{+}& \left[ \frac{\widehat{p_\mu} \widehat{S_\nu}
+ \widehat{S_\mu} \widehat{p_\nu}}{2}-
S.q \frac{\widehat{p_\mu} \widehat{p_\nu}}{(p.q)} \right]
\frac{g_3^i(x,Q^2)}{p.q}\nn\\
&+&
S.q \frac{\widehat{p_\mu}\widehat{p_\nu}}{(p.q)^2}
g_4^i(x,Q^2)+
(-g_{\mu\nu}+\frac{q_\mu q_\nu}{q^2})\frac{(S.q)}{p.q} g_5^i(x,Q^2),
\label{eqz4}
\ea
with  $ab \equiv i$
and
\be
\widehat{p_\mu} = p_\mu-\frac{p.q}{q^2} q_{\mu},~~~~~~\widehat{S_\mu}
= S_\mu-\frac{S.q}{q^2} q_{\mu}.
\label{eqz5}
\ee
Here $x = Q^2/2p.q \equiv Q^2/2M\nu$ and $Q^2 = -q^2$ is the transfered
four
momentum  squared.
$p$ and $S$ denote the four vectors of the
nucleon momentum and spin, respectively, with
$ S^2=-M^2$ and, $S.p = 0 $.
$g_{V_i}$ and $g_{A_i}$ are the vector and axialvector
couplings of the bosons exchanged in the respective subprocesses.
For charged current interactions
$U_{ff^\prime}$ denotes the
Cabibbo-Kobayashi-Maskawa matrix.
The hadronic tensor (\ref{eqz4})
was constructed using both Lorentz and time reversal
invariance and current conservation.

In previous analyses
partly different notations for the hadronic tensor
have been used. To allow for  direct comparisons
with earlier results
we relate the definition
of structure functions given in eq.~(\ref{eqz4}) to that of other
authors in table~1 for convenience~\footnote{A more comprehensive
comparison is given in~\cite{BK}.}.

\begin{center}
\begin{tabular}{||c||c|c|c|c||}\hline \hline
{\sf our notation}      & $\ct{a1}$
    & $\ct{a3}$ & $\ct{a7}$ & $\ct{a77}$\\ \hline \hline
$g_1$ & $g_1$ &  $g_1$      & $g_1$     & $g_1$\\
$g_2$ & $g_2$ &  $g_2$      & $g_2$  & $g_2$\\
$g_3$ & $-g_3$ &  $
(g_4-g_5)/2$      & $b_1+b_2$  & $(A_2-A_3)/2$ \\
$g_4 $
& $g_4-g_3$
& $g_4$
&$a_2+b_1+b_2$ &$A_2$
\\
$g_5$
& $-g_5$
&$ g_3$
&$a_1$ & $A_1$

\\  \hline
\end{tabular}
\end{center}
\small

\vspace{3mm}
\noindent
\small
{\sf Table~1:}~The definition of polarized deep inelastic scattering
structure functions in different conventions.\footnote{Note that
in part of the
above papers only the structure functions being related to longitudinal
nucleon polarization were delt with.}
\normalsize

\vspace{2mm}
%
\section{Operator Product Expansion}
\label{sect3}
%
The forward Compton amplitude, $T_{\mu\nu}^{ij}$,
is related to the hadronic tensor
by
\be
W_{\mu\nu}^{ij} = \frac{1}{2\pi} Im T^{ij}_{\mu\nu}
\label{eqB1}
\ee
where
\be
T_{\mu\nu}^{ij}
=i\int d^4xe^{iqx}\langle pS\mid(T{J_\mu^i}^\dagger (x)J_\nu^j(0))\mid
pS\rangle.
\label{eqB2}
\ee
It may be represented in terms of the amplitudes
$\left . T_k^i(q^2, \nu) \right|_{k=1}^3$ and
$\left . A_k^i(q^2, \nu) \right|_{k=1}^5$ analogously to (\ref{eqz4})
substituting
\be
F_1 \rightarrow T_1,~~~~~~~F_{2,3} \rightarrow \frac{p.q}{M^2} T_{2,3}
\label{eqB3}
\ee
and
\be
g_{1,5} \rightarrow
\frac{p.q}{M^2} A_{1,5},~~~~~~~g_{2,3,4}  \rightarrow
\frac{(p.q)^2}{M^4} A_{2,3,4}.
\label{eqB3A}
\ee
Near the light cone the forward Compton amplitude has the representation
\ba
T^{ab}_{\mu\nu, ij}   &=&
   \frac{2i}{(2\pi)^2(x^2 - i0)^2}  \left [
\bar q(x)\gamma_\mu(g_{V_i}+g_{A_i}
\gamma_5)
\not \!{x}
\gamma_\nu(g_{V_j}+
g_{A_j}\gamma_5)\lambda^a\lambda^b q(0) \right.
\nn\\
&{-}&~~~~~~~~~~~~~~~~~~~~ \left .
\bar q(0)\gamma_\nu(g_{V_j}+g_{A_j}\gamma_5)
\not \!{x}
\gamma_\mu(g_{V_i}+
g_{A_i}\gamma_5)\lambda^b\lambda^a q(x) \right ],
\label{eqB4}
\ea
where  $\lambda^a$ denote the $SU(N_f)$ matrices.
The spin dependent part of $T_{\mu\nu, ij}^{ab}$ is
\be
\begin{array}{cl}
T_{\mu\nu, ij}^{spin, ab} &
=  \frac{\displaystyle
2 x^\alpha}{\displaystyle
(2\pi)^2(x^2-i0)^2}  \\
\times &
 \left \{
(g_{V_1}g_{V_2}+g_{A_1}g_{A_2}) \epa  \left [
                        (if^{abc}+\widetilde{d}^{abc})
\bq(x)\gb\g\lc q(0)
                       -(if^{abc}-\widetilde{d}^{abc})
\bq(0)\gb\g\lc q(x) \right ] \right.
 \\
 +& \left. (g_{V_1}g_{A_2}+g_{A_1}g_{V_2})
                       \Sa \left[
 (if^{abc} + \widetilde{d}^{abc})\bq(x)\gb\g\lc q(0)
              + (if^{abc} - \widetilde{d}^{abc})
\bq(0)\gb\g\lc q(x) \right ] \right \},\\
\end{array}
\label{eqB5}
\ee
with
\be
S_{\mu \alpha \nu \beta} = g_{\mu \alpha} g_{\nu \beta}
                         + g_{\mu \beta } g_{\nu \alpha}
                         - g_{\mu \nu   } g_{\alpha \beta}
\ee
and
\be
\widetilde{d}^{abc} \lambda_c = \frac{2}{N_f} \delta^{ab} +
d^{abd} \lambda_c.
\ee
We further represent (\ref{eqB5}) in terms of a Taylor series around
$x=0$. The amplitudes $A_k(q^2, \nu)|_{k=1}^5$ can be related
to the expectation values of
a symmetric and an
antisymmetric operator
emerging in the Taylor expansion,
$\langle pS|\Theta_{S,A}^{\beta
\left\{\mu_1 ... \mu_n\right\}}|pS \rangle$,
and obey the following crossing relations:
\ba
A_{1,3}(q^2, -\nu) &=& ~A_{1,3}(q^2, \nu) \label{eqXMP}
\\
A_{2,4,5}(q^2, -\nu) &=& -A_{2,4,5}(q^2, \nu)
\label{eqAMP}
\ea
for neutral current interactions.
One finally obtains the following expressions for the
moments of structure functions
using standard techniques.
\ba
\int_0^1 dx x^n g_1^j(x,Q^2) &=& \frac{1}{4}
\sum_q \alpha_j^q a_n^q,~~~{\ } n=0,2...,
\label{eqYMP}
\\
\int_0^1 dx x^n g_2^j(x,Q^2) &=&  \frac{1}{4}
\sum_q \alpha_j^q
      \frac{n (d_n^q -a_n^q)}{n + 1} ,~~~{\ } n=2,4...,   \\
\int_0^1 dx x^n g_3^j(x,Q^2) &=&
 \sum_q  \beta_j^q
      \frac{a_{n+1}^q}{n + 2} ,~~~{\ } n=0,2...,   \\
\int_0^1 dx x^n g_4^j(x,Q^2) &=&
\frac{1}{2} 
  \sum_q 
\beta_j^q
      a_{n+1}^q ,~~~{\ } n=2,4...,   \\
\int_0^1 dx x^n g_5^j(x,Q^2) &=&
\frac{1}{4} 
  \sum_q 
\beta_j^q
      a_{n}^q ,~~~{\ } n=1,3...~~.
\label{eqg1M}
\ea
Here we adopt
the notation of~\cite{RLJ} and
$a_n^q$ and $d_n^q$ are the matrix elements which are
related to the expectation
values of
$\langle pS|\Theta_{S}^{\beta\left\{
\mu_1 ... \mu_n\right\}}|pS \rangle$
and
$\langle pS|\Theta_{A}^{\beta\left\{
\mu_1] ... \mu_n\right\}}|pS \rangle$,
respectively. The factors $\alpha_j^q$ and $\beta_j^q$ are given by
\ba
\left(
\alpha_{|\gamma|^2}^q, \alpha_{|\gamma Z|}^q, \alpha_{|Z|^2}^q \right)
&=& \left [
e_q^2, 2 e_q g_V^q, (g_V^q)^2 + (g_A^q)^2 \right ] \\
\left(
\beta_{|\gamma Z|}^q, \beta_{|Z|^2}^q \right)
&=& \left [
2 e_q g_V^q,
2 g_V^q g_A^q \right ]
\label{eqAlBe}
\ea
Analogous relations to (\ref{eqXMP}--\ref{eqg1M})
 are derived for the charged current structure
functions~( cf.~\cite{BK}).

As well--known, the structure function $g_2(x,Q^2)$ contains also
twist--3 contributions corresponding to the matrix elements $d_n^q$.
On the other hand, all the  remaining structure functions are {\it not}
related to $d_n^q$, and contain at
lowest twist
contributions  of twist--2 only.
We will disregard the terms $d_n^q$  in the
subseqent discussion.
The twist--2 contributions are
related by the equations:
\ba
g_2^i(x) &=& -g_1^i(x) + \int_x^1\frac{dy}{y}g_1^i(x),
\label{qq7}  \\
g_4^j(x) &=& 2xg_5^j(x),
\label{qq8} \\
g_3^j(x)&=&4x\int_x^1\frac{dy}{y}g_5^j(y),
\label{qq9}
\ea
where $i=\gamma,\gamma Z, Z, W $ and $j=\gamma Z, Z, W $.
Eqs.~(\ref{qq7}) and (\ref{qq8}) are the Wandzura--Wilczek~\cite{WW}
and Dicus~\cite{DIC} relations, and eq.~(\ref{qq9}) is a {\it new}
relation. 

Recently the first two moments of $g_3$
were calculated in~\cite{FRANKF}. They agree with our general relation
eq.~(\ref{qq9}). We do not confirm a corresponding relation for the
structure function $A_3$~(cf.~table~1)
given
in~\cite{a77} previously, which also
disagrees with the lowest moments
given in~\cite{FRANKF}.

Eqs.~(\ref{qq8},\ref{qq9}) yield the sum rules
\be
\int_0^1 dx x^n \left [ g_3^k(x,Q^2) - \frac{2}{n+2} g_4^k(x,Q^2)
\right] = 0.
\label{qq10}
\ee
For $n = 0$ one obtains
\be
\int_0^1 dx   g_3^k(x,Q^2)  =
\int_0^1 dx   g_4^k(x,Q^2).
\label{qq11}
\ee

Two of the five spin--dependent structure functions $\left. g_k^j\right
|_{k=1}^5$ are linearly independent. We will express the remaining ones
using $g_1^j$ and $g_5^j$ as a basis given by
\ba
g_1^j(x,Q^2)  &=&
\frac{1}{2} \sum_q \alpha_j^q \left [ \Delta q(x,Q^2)
+ \Delta \overline{q}(x,Q^2) \right ],\\
g_5^j(x,Q^2) &=&
\frac{1}{2} \sum_q \beta_j^q
\left [\Delta q(x,Q^2)  - \Delta \overline{q}(x, Q^2) \right ]
\ea
for the neutral current reactions. For charged current $lN$ scattering
one obtains:
\ba
g_1^{W^{-(+)}}(x,Q^2) &=&
\sum_q \left [\Delta q_{u(d)}(x,Q^2)
+ \Delta \overline{q}_{d(u)}(x, Q^2)
\right ],
\\
g_5^{W^{-(+)}}(x,Q^2) &=& - \sum_q
\left [\Delta q_{u(d)}(x,Q^2)
- \Delta \overline{q}_{d(u)}(x, Q^2)
\right ].
\ea

In figure~1 the behaviour of the twist--2 contributions to the
structure functions
$\left. g_k^j(x,Q^2) \right|_{k=1}^j$ are compared
for
$j = |\gamma|^2, |\gamma Z|$ and $|W^-|^2$ in leading order QCD
for the range $10^{-4} < x$  and $10 \GeV^2 \leq Q^2 \leq 10^4 \GeV^2$.
Here we refer to the
parametrization~\cite{GRVS} of the parton densities as one example.
Whereas  the absolute values of the
structure functions $g_{1,2,5}(x,Q^2)$ grow for $x \rightarrow 0$
$g_{3}^j$ and $g_4^j$ are predicted to vanish as
$x \rightarrow 0$. 
In the parametrization~\cite{GRVS} the structure functions $g_3^j$ to
$g_5^j$
are found to be positive for $j = \gamma Z$ and  negative for
$j = W^-$, while $g_1$ takes
negative values for $x \lsim 10^{-3}... 3 \cdot 10^{-4}$. For larger
values of $Q^2$ the twist--2 contribution to $g_2^k$ is predicted to be
positive, while for some current combinations it can take negative
values in the small $x$ region again.

Currently the experimental data on $g_1^n$ and
$g_1^p$ constrain the parton densities $\Delta q$ and
$\Delta \overline{q}$ in the kinematical range $10^{-2} \lsim x$
and the predictions for the small $x$ range result from extrapolations
only. Other parametrizations (see~\cite{LADIN} for a recent compilation)
agree in the range of the current data but differ in size in the
range of small~$x$. Clearly
more data, particularly in the low $x$ region, are
needed to yield better constraints on the flavour structure of
polarized structure functions.

%
\section{Covariant Parton Model}
\label{sect4}
%
In the covariant parton model the hadronic
tensor for deep inelastic scattering is given by
\be
W_{\mu\nu, ab}(q,p,S)=\sum_{\lambda, i} \int d^4k
f_{\lambda}^{q_i}(p,k,S)
w_{\mu\nu ,ab, \lambda}^{q_i}(k,q) \delta[(k+q)^2-m^2].
\label{eq1}
\ee
Here $w_{\mu\nu, ab, \lambda}^{q_i}(k,q)$
denotes the hadronic tensor at the
quark level,
$f_{\lambda}^{q_i}(p,k,S)$ describes
the quark and antiquark
distributions  of the hadron,
$\lambda$ is the quark helicity,  $k$ the virtuality
of the initial state parton, and $m$ is the quark mass.

The spin-dependent part of the hadronic tensor at the
quark level takes the
following form:
\ba
w_{\mu\nu, ab, \lambda}^{q_i, spin}(k,q)
&=&\lambda \left\{
2i\epa [g_{A_a}^{q_i}g_{A_b}^{q_i}k_\alpha
n_\beta+(g_{A_a}^{q_i}g_{A_b}^{q_i}+
g_{V_a}^{q_i}g_{V_b}^{q_i})q_\alpha n_\beta] \right.
\nn\\
&+& \left.
g_{V_a}^{q_i}g_{A_b}^{q_i}[2k_\mu n_\nu-(n.q)g_{\mu\nu}]+
g_{A_a}^{q_i}g_{V_b}^{q_i}[2n_\mu k_\nu-(n.q)g_{\mu\nu}]
\right \},
\label{eq2}
\ea
where $n$ is the spin vector of the
off-shell parton~\cite{a17}
\be
n_\sigma=\frac{m   p.k}{\sqrt{(p.k)^2 k^2 - M^2 k^4}}(k_\sigma -
\frac{k^2}{p.k} p_\sigma).
\label{eq3}
\ee
For massless quarks the spin dependent quark densities
$\Delta f^{q_i}  = f_+^{q_i} - f_-^{q_i}$ obey, due to covariance
(cf. \cite{a18}),
\be
\Delta f^{q_i}
(p.k,S.k,k^2) =  - \frac{S.k}{M^2} \hat{f}^{q_i}(p.k,k^2).
\label{eq43}
\ee
We further decompose  the spin dependent part of the hadronic
tensor $W_{\mu\nu}$ into a longitudinal and a transverse component
with respect to the nucleon spin $S^{\mu}_{\parallel} = p^{\mu}
+ {\cal O}(M^2/\nu)$ and $S^{\mu}_{\perp} = M(0,1,0,0)$ in the infinite
momentum frame $p = (\sqrt{M^2 + \pvec^2},0,0,\pvec)$:
\ba
W_{\mu\nu}^{j,\|}
=i\epa \frac{q_\alpha p_\beta}{\nu}g_1^j(x)+\frac{p_\mu
  p_\nu}{\nu}g_4^j(x)-g_{\mu\nu}g_5^j(x) ,\nn\\
W_{\mu\nu}^{j, \bot}=
i\epa \frac{q_\alpha S_\beta^\bot}{\nu} \left [
 g_1^j(x)+g_2^j(x) \right ]
+\frac{p_\mu S_\nu^\bot+p_\nu S_\mu^\bot}{2\nu}g_3^j(x).
\label{eq4}
\ea
with $j \equiv ab$.
Using the Sudakov representation for
\be
k = xp + \frac{k^2 + k_{\perp}^2 - x^2 M^2}{2 x \nu} (q + xp) + k_{\perp}
\label{eq5}
\ee
the structure functions $g_k^j(x)$
are obtained from (\ref{eq1},\ref{eq2}) in the Bjorken limit
$Q^2, \nu \rightarrow \infty,$\newline
$x = const.$ by
\ba
g_1^j(x)&=&\frac{\pi xM^2}{8}
\sum_q \alpha_q^j
\int_x^1dy(2x-y)
\widehat{h}_{q}(y) ,\nn\\
g_2^j(x)&=&
\frac{\pi xM^2}{8}
\sum_q \alpha_q^j
\int_x^1dy(2y-3x)
\widehat{h}_{q}(y) ,\nn\\
g_3^j(x)&=&
\frac{\pi x^2M^2}{2}
\sum_q \beta_q^j
\int_x^1dy(y-x)
\widehat{h}_{q}(y) ,\nn\\
g_4^j(x)
&=&\frac{\pi x^2M^2}{4}
\sum_q \beta_q^j
\int_x^1dy(2x-y)
\widehat{h}_{q}(y) ,\nn\\
g_5^j(x)
&=&\frac{\pi xM^2}{8}
\sum_q \beta_q^j
\int_x^1dy(2x-y)
\widehat{h}_{q}(y).
\label{eq6}
\ea
for neutral current interactions,
where
$ y=x+k^2_\bot/(xM^2)$ and
$\widehat{h}_{q}(y)=\int dk^2 \hat{f}_{q}(y,k^2)$.
The corresponding relations for
charged current scattering are given in~\cite{BK}.
The expressions
for $g_1^{em}$ and $g_2^{em}$
have been obtained in \cite{a17,a18}
already.

Again the structure functions given in eqs.~(\ref{eq6}) may be
expressed in terms
of two independent
structure functions in lowest order QCD:
\ba
g_2^i(x)&=& -g_1^i(x) + \int_x^1\frac{dy}{y}g_1^i(y),
\label{eq7}   \\
g_4^j(x)&=& 2x g_5^j(x),
\label{eq8}  \\
g_3^j(x)&=&4x\int_x^1\frac{dy}{y}g_5^j(y).
\label{eq9}
\ea
These relations agree with those found using the operator product
expansion in section~2, eqs.~(\ref{qq7}--\ref{qq9}).

As examples
one may derive from (\ref{eqYMP}--\ref{eqg1M},\ref{eq6}) the relations
\be
  \left [ g_3^{\nu n}(x,Q^2) - g_3^{\nu p}(x,Q^2) \right ]
=
12x \left [ \left ( g_1(x,Q^2)
                  + g_2(x,Q^2) \right )^{ep}
          - \left ( g_1(x,Q^2)
                  + g_2(x,Q^2) \right )^{en}  \right ]^{|\gamma|^2}
\label{DIC1}
\ee
\ba
6x \left [         g_2^{en}(x,Q^2)
                  - g_2^{ep}(x,Q^2) \right ]^{|\gamma|^2}
=  \left [ \left ( g_4(x,Q^2)
                  - \frac{g_3(x,Q^2)}{2} \right )^{\nu n}
         - \left ( g_4(x,Q^2)
                  - \frac{g_3(x,Q^2)}{2} \right )^{\nu p}
 \right ]. \nonumber\\
\label{DIC2}
\ea
Eqs.~(\ref{DIC1})
 and (\ref{DIC1}, \ref{DIC2}) were
given first in~\cite{a1B} and \cite{DIC}, respectively.
In a similar way various other relations follow
for other current combinations.

Let us now derive the light quark mass corrections to the structure
functions $g_j(x)\left|_{j=1}^5 \right.$.
We follow the treatment of ref.~\cite{a18} where it was shown that
as in the massless case the polarized structure functions can be
expressed in terms of  functions $\tilde{h}_q(y, \rho)$, with
$\rho = m^2/M^2$,
and the corresponding perturbative coefficient functions.
Due to the flavour dependence of the couplings, $g_{V_i}$ and $g_{A_i}$,
one has in general to introduce the functions
$\tilde{h}_{q}(x, \rho)$
even if the ratio of $m/M$ is treated as a single parameter.
In most of the cases given below a {\it single} function,
however, suffices for an {\it effective}
 parametrization.
The  mass dependent structure functions are given by:
\ba
g_1^j(x,\rho) &=&
\frac{\pi M^2x}{8} \sum_q \alpha_q^j
\int_{x+\frac{\rho}{x}}^{1+\rho}dy
\left [ x(2x-y) + 2 \rho \right ]
\tilde{h}_{q}(y,\rho),
\label{eqAA}
\\
g_2^j(x,\rho) &=&
\frac{\pi M^2}{8} \sum_q \alpha_q^j
\int_{x+\frac{\rho}{x}}^{1+\rho}dy
\left [x (2y-3x) -  \rho \right ]
\tilde{h}_{q}(y,\rho)
-
\frac{\pi m^2}{4} \sum_q \gamma_q^j
\int_{x+\frac{\rho}{x}}^{1+\rho}dy
\tilde{h}_{q}(y,\rho),
\label{eqAC}
\label{eqVIO}
\\
g_3^j(x,\rho) &=& \frac{\pi M^2 x^2}{2}   \sum_q  \beta_q^j
\int_{x+\frac{\rho}{x}}^{1+\rho}dy
(y-x) \tilde{h}_{q}(y, \rho),
\\
g_4^j(x,\rho) &=& 2x g_5(x),
\label{eqAD}
\\
g_5^j(x,\rho) &=&
\frac{\pi M^2}{8} \sum_q \beta_q^j
\int_{x+\frac{\rho}{x}}^{1+\rho}dy
\left [x (2x-y) + 2 \rho \right ] \tilde{h}_q(y,\rho),
\label{eqAB}
\ea
with $\gamma_q^j = g_{A_a}^q g_{A_b}^q, j~\equiv~ab$, and
\be
\tilde{h}_{q}(y, \rho)
=  \int dk^2 \hat{f}_{q}(y,k^2, \rho).
\label{eq11}
\ee
Corresponding relations are obtained for charged current scattering.
The last definition applies a slightly different convention than used in
ref.~\cite{a18}.
The functions $\tilde{h}_{q}(y, \rho)$ can be determined
from the different measured
structure functions
in phenomenological analyses.
In the case of the non--photonic structure functions
the direct determination of $\tilde{h}_q$
is complicated due to the fact that these structure functions are
difficult to unfold from the measured scattering cross sections.
However, one may still use the relations~(\ref{eqAA}--\ref{eqAB})
 in  global
analyses of polarization asymmetries at large $Q^2$ as corrections
in the determination of $g_1(x,Q^2)$ in this kinematical range.

The relations (\ref{eqAA}) and (\ref{eqAC}) agree with those derived
in ref.~\cite{a18} recently for photon exchange, where $\gamma_q^j
 = 0$.
Note that for the contributions due to $Z$ or $W^{\pm}$ exchange
a new contribution to $g_2 \propto m^2/M^2$ emerges.
In a different context similar terms were obtained in \cite{a1}
as the only contributions to $g_2^B$, $B=Z, W^{\pm}$.
The
Burkhardt--Cottingham sum rule
\begin{equation}
\int_0^1 dx g_2^k(x) = 0
\end{equation}
is valid for $\rho \neq 0$
iff $\gamma_{q}^j
= g_{A_a}^{q_i} g_{A_b}^{q_i} = 0$, i.e.
for pure $Z$ and $W^{\pm}$
it is violated due to the second term in
(\ref{eqVIO}).
For charged current interactions,
on the other hand,
\begin{equation}
\int_0^1 dx~x \left [  g_1^k(x)+ 2 g_2^k(x) \right ] = 0
\end{equation}
is valid  for all values of $\rho$.
The sum--rule eq.~({\ref{qq11})
\begin{equation}
\int_0^1 dx g_3^k(x, \rho) = \int_0^1 dx g_4^k(x, \rho)
\end{equation}
holds also
for massive quarks.
Finally
also  the  Dicus relation
between $g_4(x)$ and $g_5(x)$~(\ref{eqAD}) obtains no quark mass 
corrections.

An illustration of the relative
size of the mass terms for the different
structure functions is given in figure~2 for $m/M = 0.005$~and~0.010
in terms of  relative correction factors.
These mass values mark the typical range for the light current
quark masses $m_u = 5.1 \pm 0.9 \MeV$
and $m_d = 9.3 \pm 1.4 \MeV$~\cite{BERN}.

To obtain a first estimate we use the same parametrization for
all the functions $\tilde{h}(y,\rho)$~\footnote{We
would like to
thank R.G. Roberts  for providing us with the fit parameters of the
function $\tilde{h}(y,\rho)$ determined in ref.~\cite{a18}.}.
Due to the proportionality of $g_1^j(x, \rho)$, $g_4^j(x, \rho)$, and
$g_5^j(x, \rho)$ only the ratios $\left .
g_k^j(x, \rho)/g_k^j(x)\right|_{k=1}^3$ are different.
The present data constrain these ratios to a range around unity
in the region of $x \gsim 0.02$~\footnote{ The spike in
$g_2^j(x, \rho)/g_2^j(x)$ is due to
the zero in $g_2(x)$.}. The ratio for $g_3^j$ is somewhat closer
to unity than that for $g_1$ and $g_2$ at lower $x$ values.
At smaller values of $x$
and larger values of $\rho$
the correction factors differ for the various structure functions.
%
\section{Conclusions}
\label{sect6}
We have derived the twist--2 contributions to the polarized structure
functions in lowest order QCD including weak currents. The results
obtained using the operator product expansion and the covariant
parton model agree. In lowest order two out of five structure functions
are independent
for the respective current combinations
and the
remaining structure functions are related by three linear operators.
A new relation between the structure functions $g_3^j$ and $g_5^j$ was
derived. As a consequence the first moment of $g_3^j$ and $g_4^j$ are
predicted to be equal.

The light quark mass corrections to the structure functions
$\left.
g^j_k \right|_{k=1}^5$ were calculated in the covariant parton model.
The first moments of the structure functions $g_3$ and $g_4$ are
equal also in the  presence of the
quark mass corrections. The Dicus relation remains to be valid.
The Burkhardt--Cottingham sum rule is broken by a term
$\propto g_{A_a} g_{A_b} m^2/M^2$, i.e. for pure $Z$~exchange and in
charged current interactions.

\vspace{3mm}
{\bf Acknowledgement} N.K.~would like to thank DESY for the hospitality
extended to him.


\newpage
\vspace*{3cm}
\begin{center}

\vspace{-4cm}
\mbox{\epsfig{file=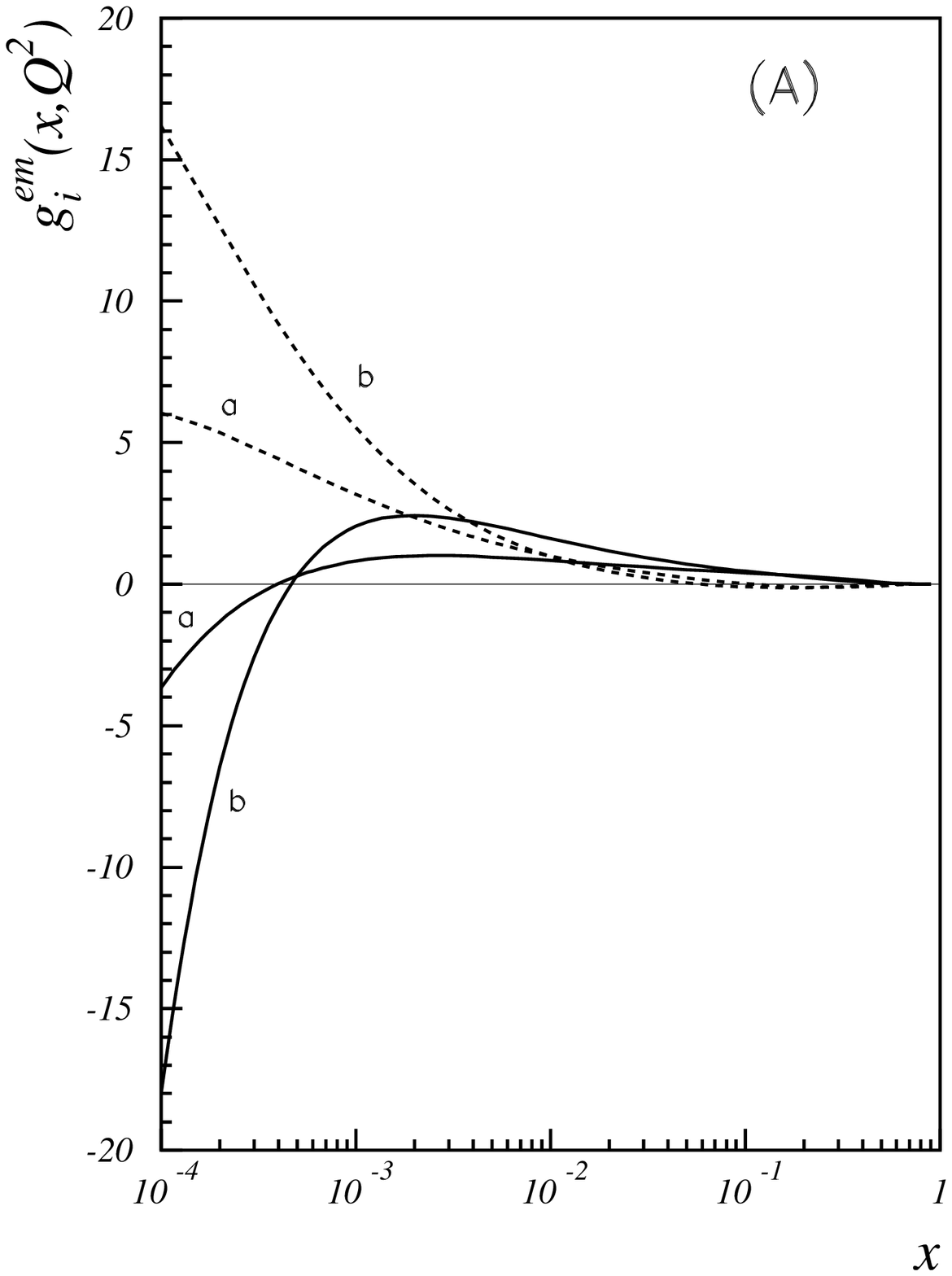,width=16cm}}
\small
\end{center}
\small
{\sf Figure~1:}~$x$ and $Q^2$ dependence of the structure functions
$\left . g^k_i(x,Q^2) \right|_{i=1}^5$ using the parton
parametrization~\cite{GRVS} (LO, STD). The lines correspond to
$Q^2 = 10 \GeV^2$~(a) and $Q^2 = 10^4  \GeV^2$~(b). The structure
functions for  photon exchange, $g_i^{em}$~(A),
$\gamma Z$ interference,
$g_i^{\gamma Z}$~(B),
and $W^-$ exchange in charged current $l N$ scattering,
$g_i^{W^-}$~(C),
are compared separately. Full lines: $g_1$, dashed lines:
$g_2$, dotted lines: $ 10 \times g_3$, dash-dotted line: $g_5$.
The structure function $g_4$ can be obtained by the
Dicus relation
$g_4 = 2x g_5$ directly.
\normalsize

\newpage
\vspace*{3cm}
\begin{center}

\mbox{\epsfig{file=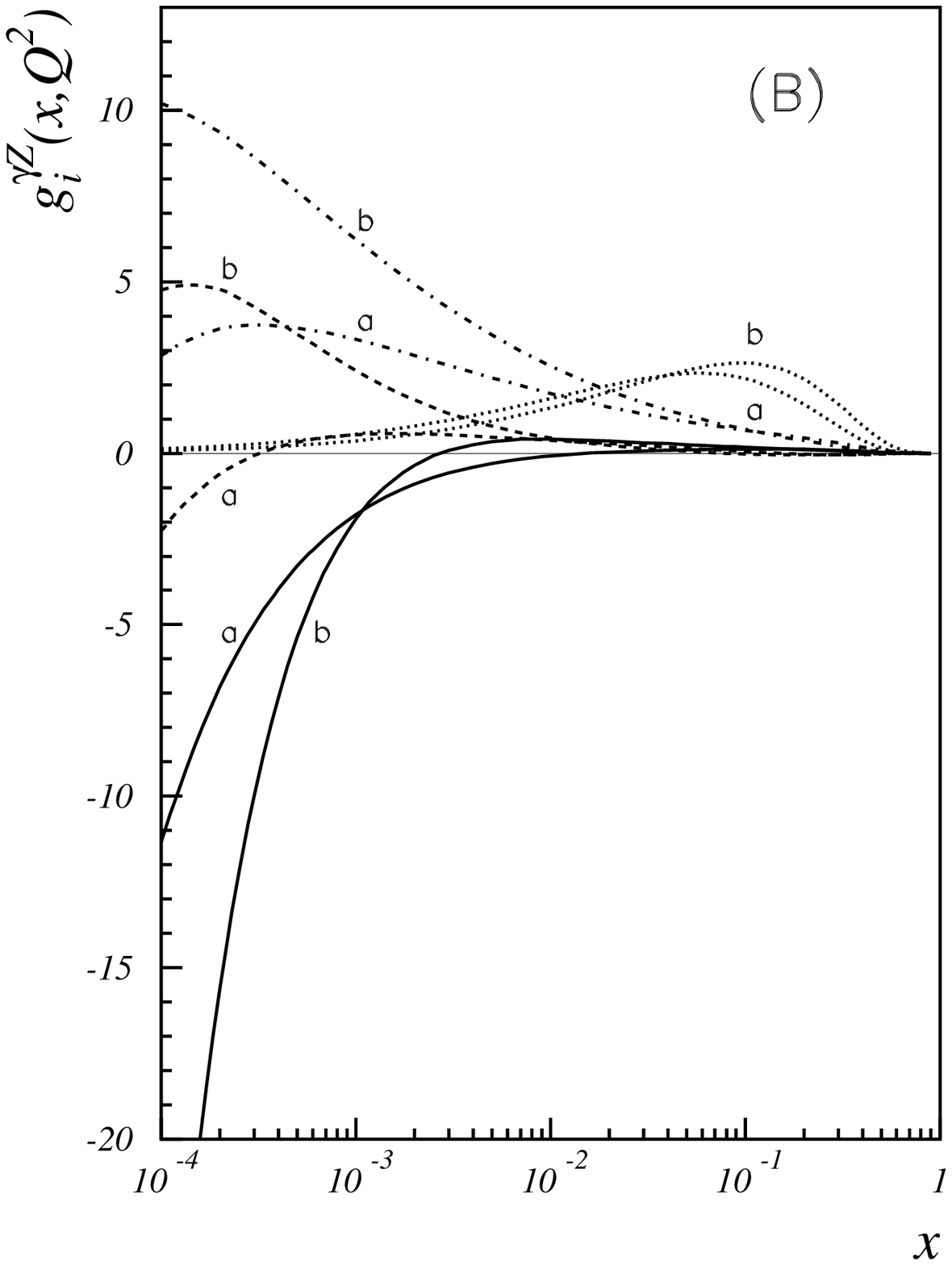,width=16cm}}
\small
\end{center}

\newpage
\vspace*{3cm}
\begin{center}

\mbox{\epsfig{file=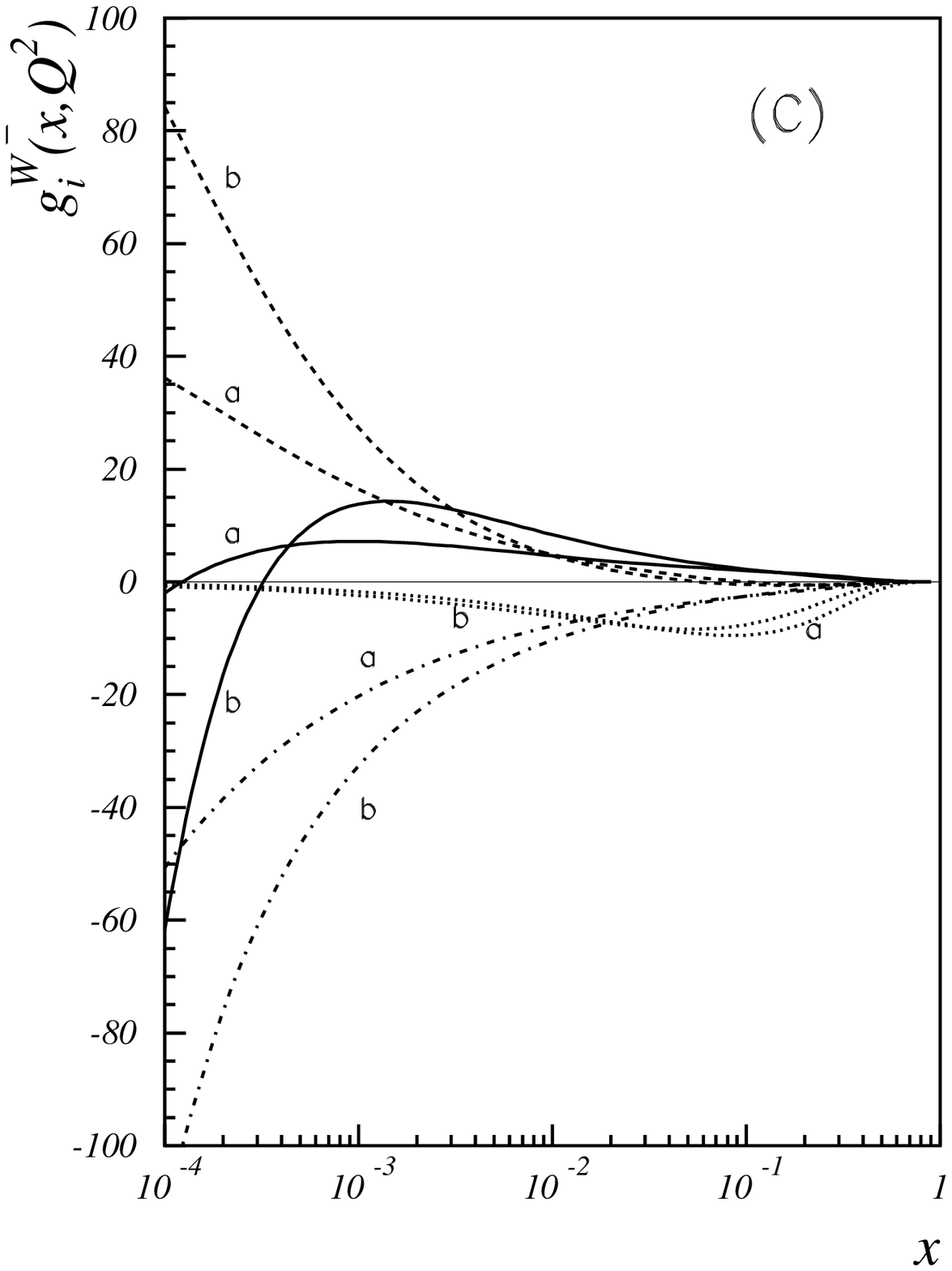,width=16cm}}
\small
\end{center}

\newpage
\vspace*{3cm}
\begin{center}

\mbox{\epsfig{file=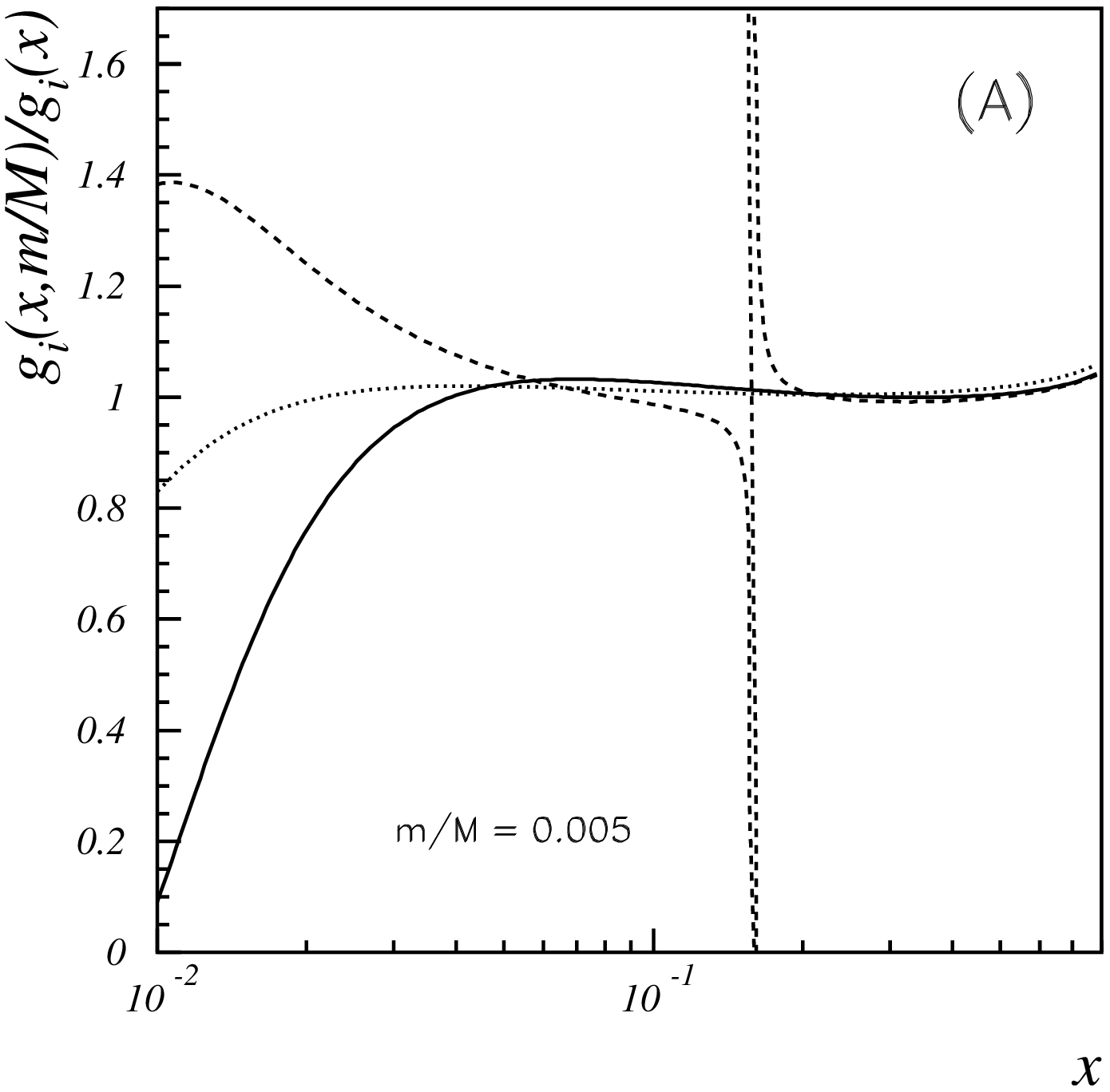,width=16cm}}
\small
\end{center}
{\sf Figure~2:}~$x$ depedence for the relative mass corrections
$g_i(x, m/M)/g_i(x, m=0)$. (A)~$m/M = 0.005$, (B)~$m/M = 0.010$.
Full lines: $g_1$, dashed lines: $g_2$
($\gamma_q^j
 = 0$, cf. eq.~(\ref{eqAC})),  dotted lines: $g_3$.
The ratios for $g_4$ and $g_5$ are identical to the ratio for $g_1$.
\normalsize

\newpage
\vspace*{3cm}
\begin{center}

\mbox{\epsfig{file=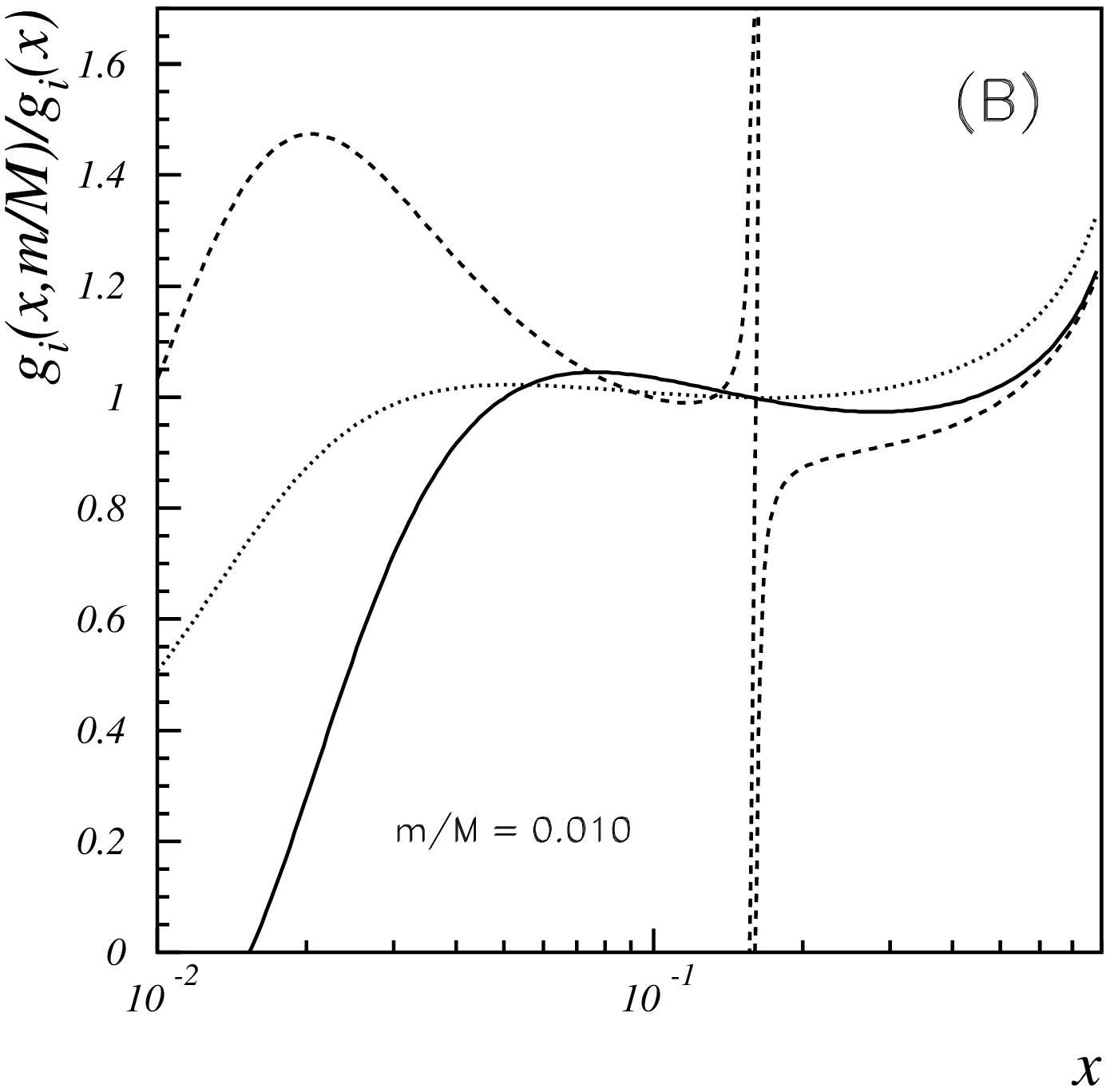,width=16cm}}
\small
\end{center}
\end{document}